\documentclass{aa}  
\usepackage{graphicx}
\graphicspath{{./}{paper_fig/}}
\usepackage{txfonts}
\usepackage{natbib}
\bibpunct{(}{)}{;}{a}{}{,}

\begin{document}

\makeatletter
\let\linenumbers\relax
\makeatother

   \title{Coronal Abundance Fractionation Linked to\\ Chromospheric Transverse MHD Waves in a Solar Active Region Observed with FISS/GST and EIS/Hinode}

   \author{Kyoung-Sun Lee\inst{1}
          \and
          Jongchul Chae\inst{1} 
          \and
          Hannah Kwak\inst{2}
          \and
          Kyuhyoun Cho\inst{3, 4}
          \and
          Kyeore Lee\inst{1}   
          \and
          Juhyung Kang\inst{1}
          \and
          Eun-Kyung Lim\inst{5}
          \and
          Donguk Song\inst{5, 6}         }

   \institute{Astronomy program, Department of Physics and Astronomy, Seoul National University, Seoul 08826, Korea\\
              \email{rolrin121@snu.ac.kr}
         \and
            Department of Astronomy and Space Science, Chungnam National University, Daejeon 34134, Korea
         \and
            Bay Area Environmental Research Institute, NASA Research Park, Moffett Field, CA 94035, USA
         \and 
            Lockheed Martin Solar and Astrophysics Laboratory, 3251 Hanover St, Palo Alto, CA 94306, USA 
         \and
            Solar and Space Weather Group, Korea Astronomy and Space Science Institute, Daejeon 34055, Korea    
         \and
            National Astronomical Observatory of Japan, 2-21-1 Osawa, Mitaka, Tokyo 181-8588, Japan 
            }

   \date{Received November 26, 2024; accepted March 16, 1997}

 
  \abstract
   {The elemental abundance in the solar corona differs from that in the photosphere, with low first ionization potential (FIP) elements showing enhanced abundances, a phenomenon known as the FIP effect. This effect is considered to be driven by ponderomotive forces associated with magnetohydrodynamic (MHD) waves, particularly incompressible transverse waves. }
   {We aim to investigate the relationship between coronal abundance fractionation and transverse MHD waves in the chromosphere. We focus on analyzing the spatial correlation between the FIP fractionation and these waves, while exploring wave properties to validate the ponderomotive force driven fractionation model.}
   {We analyze the H$\alpha$ data from the Fast Imaging Solar Spectrograph of the Goode Solar Telescope to detect chromospheric transverse MHD waves, and \ion{Si}{x} (low FIP) and \ion{S}{x} (high FIP) spectra from EUV Imaging Spectrometer onboard Hinode to determine the relative abundance in an active region. By extrapolating linear-force-free magnetic fields with Solar Dynamics Observatory/Helioseismic and Magnetic Imager magnetograms, we examine the connection between chromospheric waves and coronal composition. Around 400 wave packets were identified, and their properties, including period, velocity amplitude, propagation speed, and propagation direction, were studied.}
   {These chromospheric transverse MHD waves, mostly incompressible or weakly compressible, are found near loop footpoints, particularly in the sunspot penumbra and superpenumbral fibrils. The highly fractionated coronal region is associated with areas where these waves were detected within closed magnetic fields. Our examination of the statistics of wave properties revealed that downward-propagating low-frequency waves were particularly prominent, comprising about 43\% of the detected waves.  }
   {The correlation between abundance fractionation and transverse MHD waves, along with wave properties, supports the hypothesis that FIP fractionation occurs due to the ponderomotive force from transverse MHD waves in the chromosphere. Additionally, the observed characteristics of these chromospheric waves provide valuable observational constraints for understanding the FIP fractionation process.}

   \keywords{Magnetohydrodynamics (MHD) --
                Waves --
                Methods: observational --
                Techniques: imaging spectroscopy --
                Sun: abundances --
                Sun: atmosphere
               }
               
   \authorrunning{Lee et al.}
   \titlerunning{Coronal Abundance Fractionation Linked to Chromospheric Transverse Waves}

   \maketitle
%

\section{Introduction}

\begin{figure*}
   \centering
   \includegraphics[width=17cm]{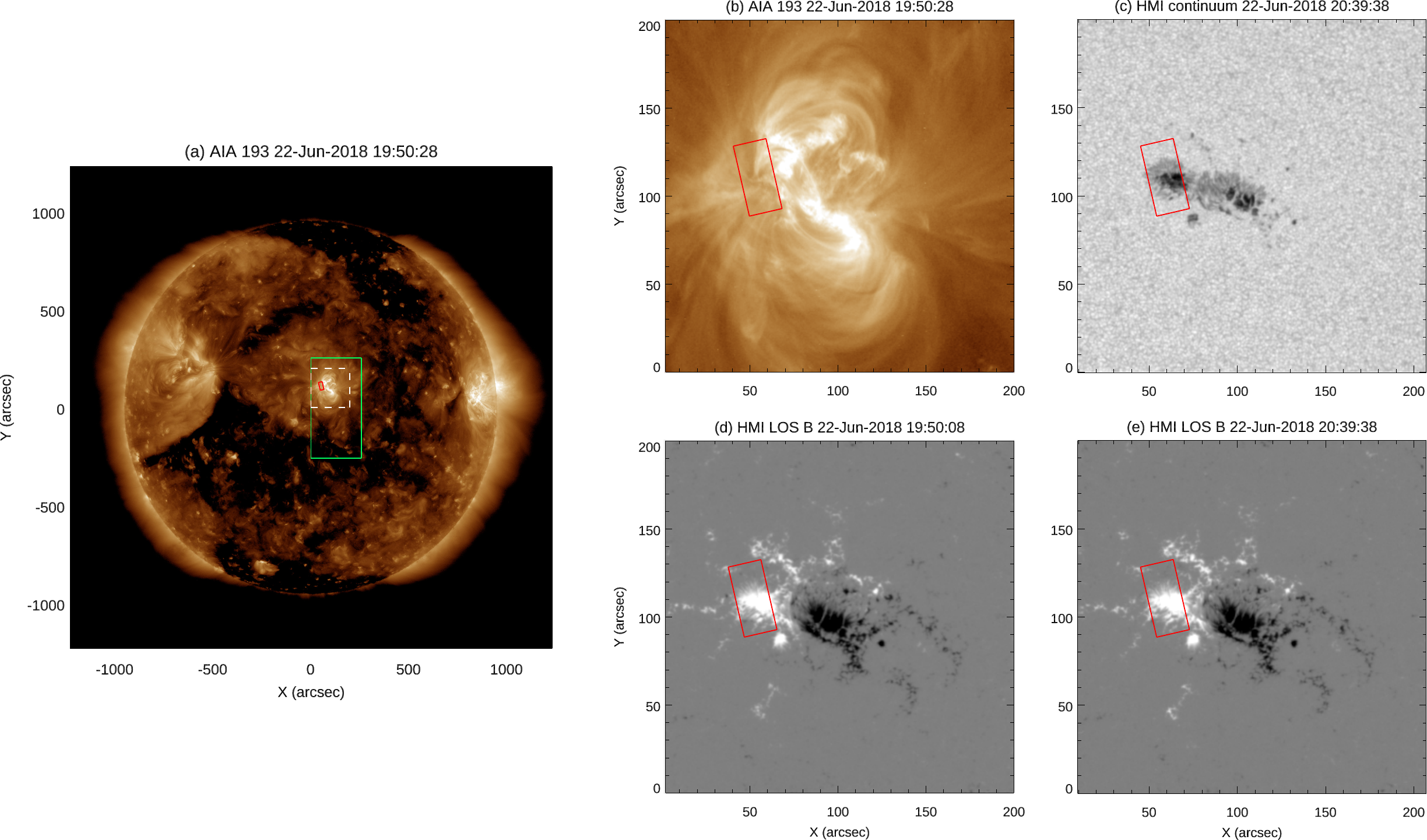}
   \caption{ Context SDO images for AR 12175 on 2018 June 22. (a) A full-sun image from SDO/AIA 193 $\mathrm{\AA}$ taken at 19:50 UT, coinciding with the EIS observation period (19:09:50 - 20:09:33 UT). Green and red boxes denote the fields of view for Hinode/EIS and FISS, respectively. The white dashed box indicates the enlarged region in the panels (b-e). The enlarged images are (b) SDO/AIA 193 $\mathrm{\AA}$ at 19:50 UT, (c) SDO/HMI continuum at 20:39 UT during FISS observation period (20:11:55 - 21:59:03 UT), (d) SDO/HMI longitudinal magnetograms at 19:50 UT, and (e) at 20:39 UT for AR 12175. }\label{fig1}
\end{figure*}
     
The elemental abundances in various regions of the solar corona differ from those in the photosphere. Typically, elements with low first ionization potential (FIP $<$ 10 eV; e.g., Si, Ar, and Fe) show anomalous enhancements (factors of 2-4) in the solar corona compared to the photosphere. In contrast, high FIP elements (FIP $>$ 10 eV; e.g., O, S, and Ca) show smaller abundance enhancements or unchanged \citep{pottasch1963, meyer1985a, meyer1985b, feldman1992, laming_etal1995, feldman_laming2000, feldman_widing2003, feldman_widing2007}. This phenomenon, known as the ``FIP effect,'' refers to the abundance fractionation between low-FIP (enhanced) and high-FIP (depleted or unaffected) elements. Based on this FIP effect, the abundance variation of the solar atmosphere is measured by the FIP bias factor, which is the ratio of the elemental abundance in the solar atmosphere to that in the photosphere. Then, the FIP bias factor of $\sim 1$ represents the photospheric plasma composition, while values between 2-4 or higher indicate coronal plasma composition fractionated by the FIP effect.

The elemental abundance fractionation depending on FIP has also been observed in the stellar coronae. Some stellar atmospheres have shown a similar FIP effect of the solar corona \citep{laming_etal1996, drake_etal1997,laming_drake1999}, while others have different abundance variations, such as no fractionation or``inverse-FIP (I-FIP) effect,'' which depletes the low FIP elements in the corona \citep{drake_etal1995, drake_etal2001, wood_etal2012}. 

With the high sensitivity and spatial resolution of EUV and X-ray spectroscopy, it has become feasible to determine spatially resolved abundances of various solar structures, including active region boundaries \citep{brooks_warren_2011, brooks_etal2015}, sunspots \citep{baker_etal2013, baker_etal2015,  baker_etal2018}, plumes and interplumes \citep{guennou_etal2015}, polar jets \citep{lee_etal2015}, and flares \citep{doschek_etal2015, doschek_warren2016, warren_etal2016, to_etal2021}. An important finding using the spatially resolved spectroscopic observations is the first detection of the I-FIP effect on the Sun \citep{doschek_etal2015, doschek_warren2016}, which was previously observed only in the late-type stars. These studies discovered the I-FIP effect in localized strong magnetic patches of flaring active regions near sunspots. Additionally, \citet{baker_etal2019, baker_etal2020} found I-FIP plasma composition in flares and light bridges, alongside the evolving magnetic field configurations.

Several models have been proposed to explain the FIP and I-FIP effect (references in the review of \citet{laming2015}). The most probable model is ``abundance fractionation by the ponderomotive force,'' which can explain both the FIP and I-FIP cases \citep{laming2004, laming2015, laming2017, laming2021}. The ponderomotive force induced by the transverse magnetohydrodynamic (MHD) waves, such as Alfv\'en waves or fast-mode waves, preferentially affects the ionized elements over neutral ones in the chromospheric plasma, which is partially ionized. Then, the force drags the low FIP elements up (or down) depending on the presence of waves, wave energy density, and their sign of gradient in the atmosphere.

In the scenario of the ponderomotive force model, transverse MHD waves, that is, Alfv\'en waves or fast-mode waves, should exist in the chromosphere for the ponderomotive force to produce the FIP or I-FIP effect. Until now, there have been only a few observational studies on the existence of Alfv\'enic perturbations related to the coronal fractionated plasma \citep{baker_etal2021, stangalini_etal2021} since it required a combination of observations from both the chromosphere and corona and was also challenging to detect or identify the 
transverse MHD waves. The studies of \citet{baker_etal2021} and \citet{stangalini_etal2021} showed that the FIP fractionated coronal structures in an active region are spatially connected to the locations where the magnetic fluctuations are dominantly detected in the chromosphere. They assumed these magnetic fluctuations were a proxy of the 
transverse MHD waves, and suggested the importance of the role of MHD waves in generating the FIP effect. 
In order to examine the hypothesis that the ponderomotive force produces the FIP/I-FIP effect, more observational evidence of the connection between the plasma composition variation and the 
transverse MHD waves with their wave properties is required.

Recent high-resolution imaging and spectroscopic observations have increasingly provided evidence of these waves \citep{Depontieu_etal2007, tomczyk_etal2007, mathioudakis_etal2013, jess_etal2015, jafarzadeh_etal2017, morton_etal2021, chae_etal2021a, chae_etal2022, kwak_etal2023}. 
Most transverse MHD waves are detected as oscillatory transverse displacements of thread-like structures (e.g., spicules, fibrils, or loops) in the solar atmosphere by using high-resolution imaging observations \citep{Depontieu_etal2007, tomczyk_etal2007, morton_etal2021}. From spectroscopic observations, transverse MHD waves were identified as the line-of-sight (LOS) velocity oscillations in the thread-like features, such as superpenumbral fibrils \citep{chae_etal2021a, chae_etal2022} and quiet region fibrils \citep{kwak_etal2023}. An advantage of the spectroscopic method is that the correlation between the velocity and temperature oscillations possibly examines whether these MHD waves are compressible or incompressible (weakly compressible). Then, the incompressible or weakly compressible waves were only identified as the 
transverse MHD waves.

In this work, we present additional observational evidence connecting coronal plasma fractionation with 
transverse MHD waves by comparing spatially resolved coronal plasma composition maps with detected transverse MHD waves in the chromosphere. To detect transverse waves, we used the spectroscopic method of \citet{chae_etal2021a, chae_etal2022}, which enables detailed analysis of their periods, velocity amplitudes, propagation speeds, and directions. Unlike previous studies, this work focuses on examining wave properties. We interpret the wave properties based on the ponderomotive force model, suggesting the possible origin of the waves and their relation to the coronal plasma fractionation.

\begin{table*}
\caption{Spectral windows in the EIS study of {HPW021VEL260x512v2}}
\label{tbl-1}
\centering
\begin{tabular}{c c c}
\hline\hline
Diagnostics &Ion  & Wavelength ($\mathrm{\AA}$) \\
\hline
  Emission Measure & \ion{Fe}{viii}   & 185.21  \\
   	& \ion{Fe}{ix}   &  188.50, 197.86  \\
   	& \ion{Fe}{x}  & 184.54  \\
   	& \ion{Fe}{xi}  & 184.79, 188.22, 188.30 \\
   	& \ion{Fe}{xii}  & 186.89, 195.12, 195.18, 203.73  \\
	& \ion{Fe}{xiii}  &  202.04, 203.83, 203.80 \\
   	& \ion{Fe}{xiv}  &  211.32, 264.79 \\
   	& \ion{Fe}{xv}  &  284.16  \\
   	& \ion{Fe}{xvi}  &  262.98  \\
  	& \ion{Fe}{xvii}  &  275.55 \\ 
  \hline
  Density & \ion{Fe}{xiii}  &  202.04, 203.83 \\
  \hline
  FIP bias & \ion{S}{x}   & 264.23  \\
  	        & \ion{Si}{x}  & 258.37  \\
 \hline	        
  \end{tabular}
\end{table*}
%


\section{Observations and analysis}

To compare the coronal plasma compositions and detected transverse MHD waves in the chromosphere, we analyzed observations of a sunspot in AR 12715 obtained on 2018 June 22. The sunspot was observed by both the Extreme-ultraviolet Imaging Spectrometer \citep[EIS,][]{culhane_etal2007} onboard Hinode \citep{kosugi_etal2007} and the Fast Imaging Solar Spectrograph \citep[FISS,][]{chae_etal2013} installed at the 1.6 meter Goode Solar Telescope \citep[GST,][]{cao_etal2010} of the Big Bear Solar Observatory. The imaging spectral data from the corona and chromosphere made it possible to investigate the relations between the coronal plasma composition and chromospheric transverse waves. For context images and alignment, we also used Atmospheric Imaging Assembly \citep[AIA,][]{lemen_etal2012} and Helioseismic and Magnetic Imager \citep[HMI,][]{schou_etal2012} onboard Solar Dynamics Observatory \citep[SDO,][]{pesnell_etal2012}. Figure~\ref{fig1} shows the context images of AR 12715 in SDO/AIA 193 $\mathrm{\AA}$, SDO/HMI continuum, and longitudinal magnetograms during EIS and FISS observing periods. The longitudinal magnetograms show negligible changes across the observation times, indicating that the sunspot remained stable without significant changes during this period. The green and red overlaid boxes indicate the position of the Hinode/EIS and FISS field of views (FOVs), respectively. The FOVs overlap in the trailing sunspot, observed nearly simultaneously in the EIS and FISS scanning rasters. 

\subsection{Hinode/EIS Observations and Analysis} 

We analyzed Hinode/EIS imaging spectral data to measure the plasma composition. The EIS observed the active region with the study {HPW021VEL260x512v2}, which ran from 19:08:50 UT to 20:09:33 UT. This EIS study is designed for observing the active region using the $2 \arcsec$ slit with a scan step size of $3 \arcsec$ for 87 scan positions, which produces a raster with FOV of $260 \arcsec \times 512 \arcsec$. The exposure time at each step is 40 seconds, and the raster scan takes about an hour. This study has 25 spectral windows and the spectral lines we used are listed in Table \ref{tbl-1}. The spectral resolution of the EIS is about 0.022 $\mathrm{\AA}$. We used the standard EIS procedure {\tt eis\_prep.pro} in the Solar SoftWare package to correct for dark current, cosmic rays, hot, warm, and dusty pixels and to remove instrumental effects of orbital variation, CCD detector offset, and slit tilt. In addition, we applied the revised radiometric calibration presented by \citet{del_zanna2013} because the response of EIS detector shows a degradation with time in the different wavelength channels after the launch of Hinode \citep{del_zanna2013, warren_etal2014}.

\begin{figure*}
   \centering
   \includegraphics[width=17cm]{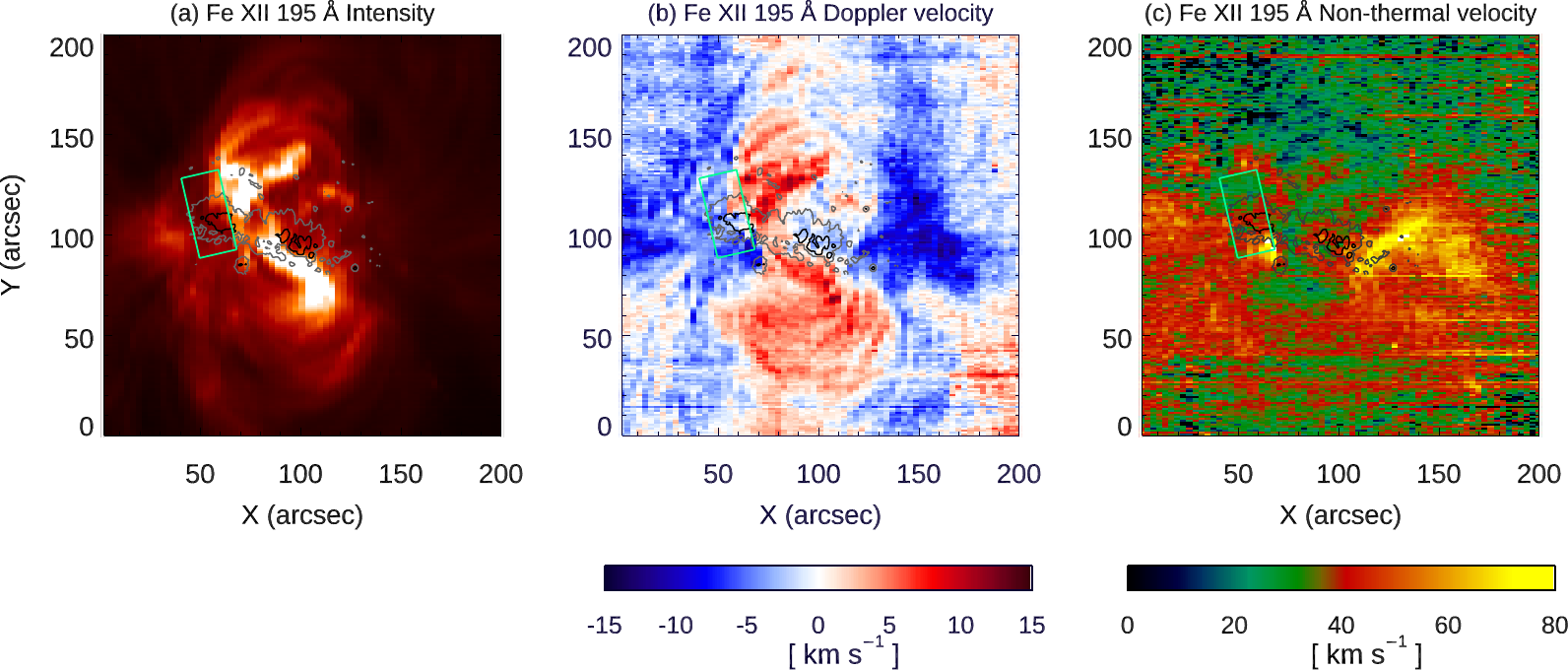}
   \caption{(a) Intensity, (b) LOS velocity, and (c) nonthermal velocity maps measured from \ion{Fe}{xii} 195.12 $\mathrm{\AA}$ from Hinode/EIS. The black and gray contours depict the umbra and penumbra boundaries, respectively. The green box indicates the FOV of FISS. The white dashed box represents the region of our interest in EIS observation.} 
   \label{fig2}
    \end{figure*}

Figure~\ref{fig2} shows the intensity, Doppler velocity, and nonthermal velocity of the \ion{Fe}{xii} 195.12 $\mathrm{\AA}$, which is the strongest spectral line among the EIS spectra formed at coronal temperature lines ($\log~(T/{\mathrm{K}})$ = 6.15). For the Doppler velocity measurements, since EIS does not have an absolute wavelength calibration, we obtain a reference wavelength by averaging the centroid wavelengths of pixels in the quieter region within the FOV. The nonthermal velocities were measured from the observed line width, which consists of thermal and nonthermal broadening combined with instrumental broadening, by assuming that the instrumental broadening of Hinode/EIS is $\sim 0.05-0.068 ~\mathrm{\AA}$ and calculating the thermal broadening from the peak formation temperature \citep{brooks&warren_2016}. The intensity map provides the context image of plasma in coronal temperature as well as is used for the co-alignment with the SDO/AIA 193 ${\mathrm{\AA}}$ filter image. Doppler velocity and nonthermal velocity maps give information on the coronal plasma dynamics in this region. 

For measuring the plasma elemental composition, we calculated the FIP bias factor from the ratio of intensity between high and low FIP elements. In this study, we used the method developed by \citet{brooks_warren_2011}, which measured the FIP bias factor from the EIS spectral line pair of \ion{Si}{x} 258.37 $\mathrm{\AA}$ (high FIP elements) and \ion{S}{x} 264.23 $\mathrm{\AA}$ (low FIP elements) having a formation temperature at around $\log~(T/\mathrm{K})$ = 6.1 - 6.2 (1.25 - 1.5 MK). To account for the temperature and density sensitivity of the line ratio, this method infers the electron density using density-sensitive line ratios. Then, it derives the differential emission measures (DEMs), which present the emission measure (EM) distribution in temperature, using Fe (low FIP elements) emissions from the consecutive ionization stages while assuming the measured densities. The details of this method can be found in \citet{brooks_warren_2011} and \citet{brooks_etal2015}. 

\begin{figure}
  \centering
  \resizebox{\hsize}{!}{\includegraphics{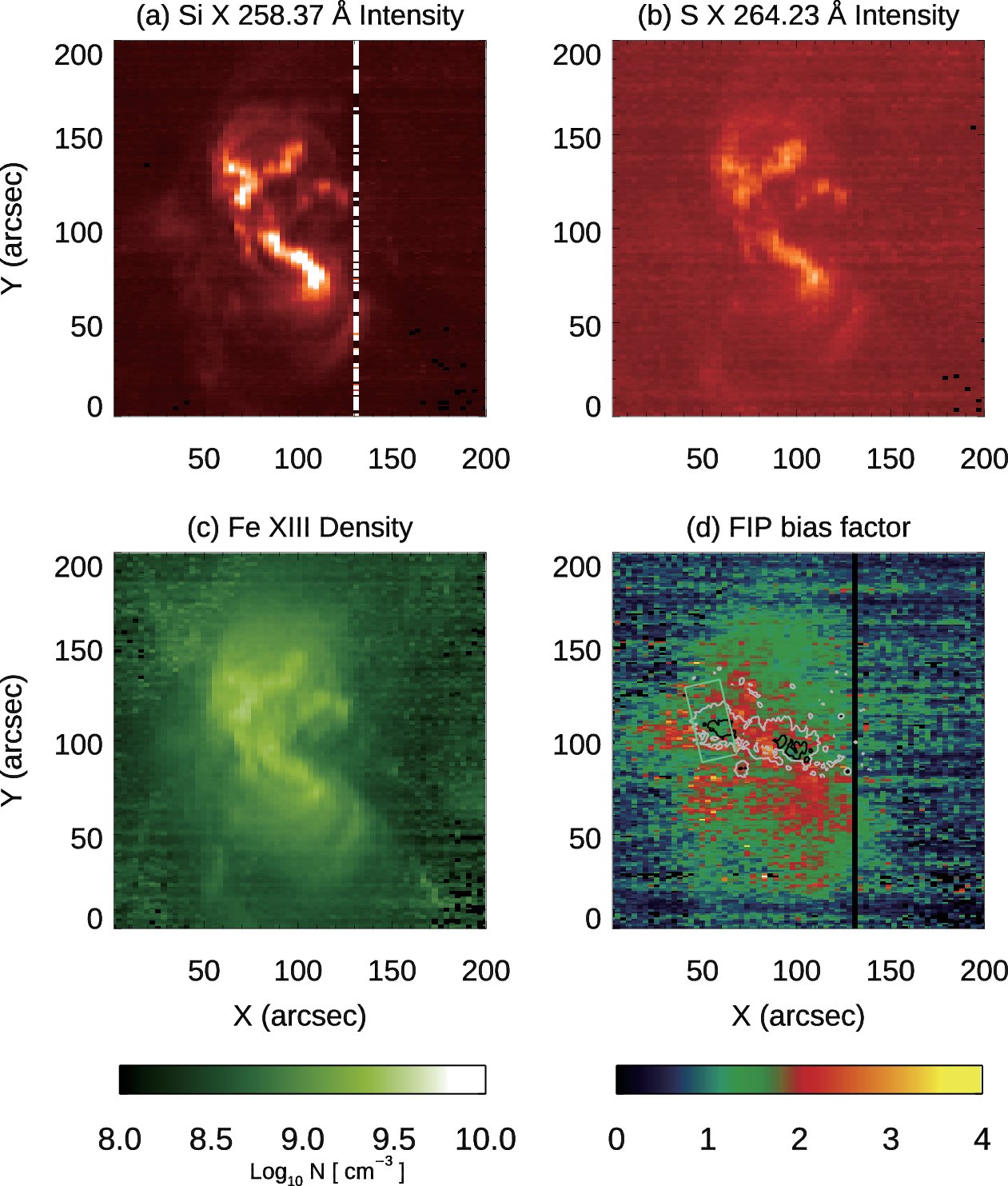}}
  \caption{Intensity maps from (a) \ion{Si}{x} 258.37 $\mathrm{\AA}$ and (b) \ion{S}{x} 264.23 $\mathrm{\AA}$. (c) Density map from the ratio of \ion{Fe}{xiii} 202.04 $\mathrm{\AA}$ and 203.83 $\mathrm{\AA}$. (d) FIP bias factor map from the ratio of \ion{S}{x} 264.23 $\mathrm{\AA}$ and \ion{Si}{x} 258.37 $\mathrm{\AA}$. The black and gray contours in panel (d) outline the umbra and penumbra boundaries, respectively. The green box corresponds to the FOV of FISS.}
  \label{fig3}
\end{figure}

Here, first, we measured line intensities of \ion{Si}{x} 258.37 $\mathrm{\AA}$, \ion{S}{x} 264.23 $\mathrm{\AA}$, density-sensitive line pair, and Fe lines for DEM measurements by fitting with single and multiple Gaussian functions. Then, we measured the density from the \ion{Fe}{xiii} 202.04 $\mathrm{\AA}$ to 203.83 $\mathrm{\AA}$ line ratio for each pixel in our interested regions. To derive the temperature distribution, we construct the DEMs for the pixels using the \ion{Fe}{viii}-\ion{Fe}{xvii} line intensities, their intensity errors, and the derived density measurements. For the error of intensities, we computed the 1$\sigma$ error from the Gaussian fit for the intensity measurement and a calibration error for the instrument. For the DEM calculation, we used the Markov-chain Monte Carlo (MCMC) method \citep{kashyap_drake1998} in the PINTofALE spectroscopy package \citep{kashyap_drake2000} and the contribution function calculated by the CHIANTI atomic database version 8.0 \citep{dere_etal1997, del_zanna_etal2015} with the ionization fractions of \citet{bryans_etal2009} and the photospheric abundances of \citet{grevesse_etal2007}. After deriving the DEM from the Fe lines, we scaled the DEM using the \ion{Si}{x} 258.37 $\mathrm{\AA}$ intensity, which is also a low FIP element. If the low FIP elements are enhanced by some factor (coronal abundance) in the observed plasma due to the FIP effect, the scaled DEM determined by both low FIP elements, Fe and Si, will be enhanced. Therefore, we estimated the intensity of the \ion{S}{x} 264.23 $\mathrm{\AA}$ line using the scaled DEM from the low FIP elements and measured the ratio of the predicted intensity to the observed intensity, which represents the FIP bias factor. Figure~\ref{fig3} shows the intensity maps of \ion{Si}{x} 258.37 $\mathrm{\AA}$ (a) and \ion{S}{x} 264.23 $\mathrm{\AA}$ (b), \ion{Fe}{xiii} density map (c), and the FIP bias factor map (d) of the region where we are interested. 

\begin{figure*}
\centering
\includegraphics[width=17cm]{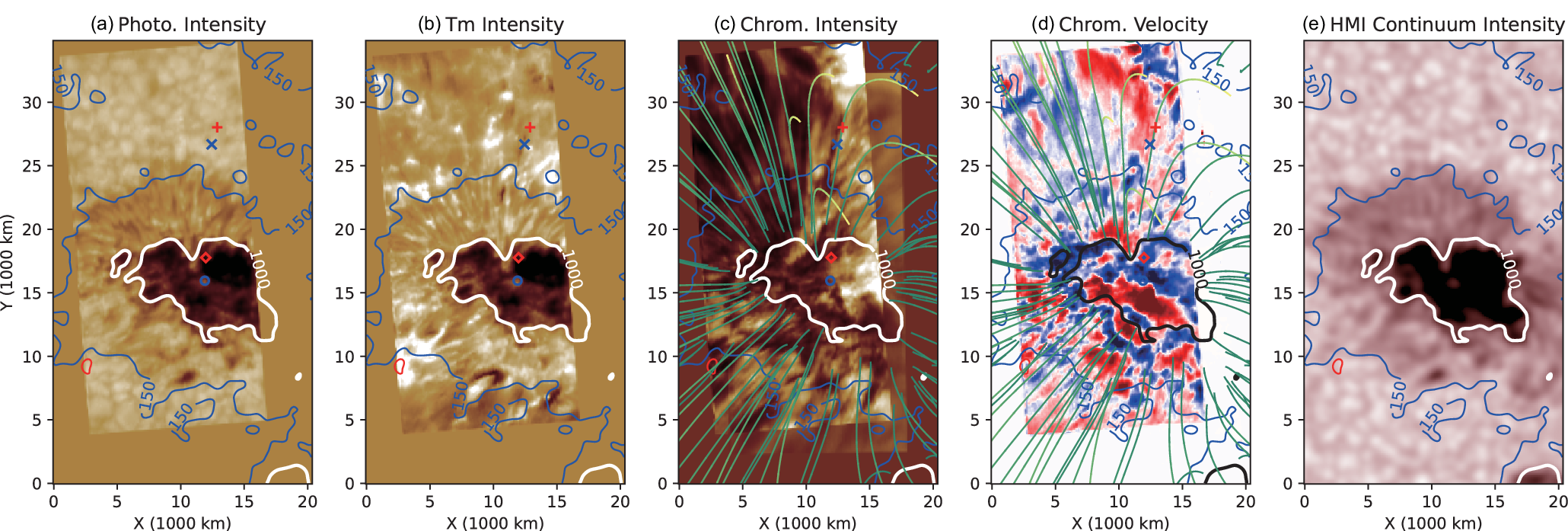} 
\caption{Maps of (a) photospheric intensity, (b) temperature minimum intensity, (c) chromospheric intensity, and (d) band-filtered (1.5 -- 4.0 minutes) chromospheric velocity taken by FISS H$\alpha$ spectra at 20:39 UT of 2018 June 22. For the alignment, (e) SDO/HMI continuum intensity map is also shown. The white and blue contours indicate positive LOS magnetogram of 1000 Gauss and 150 Gauss, repectively. The red contour marks the negative LOS magnetogram of 150 Gauss. The green lines represent the magnetic field lines constructed using the linear-force-free extrapolation method, with footpoints located at regions where the LOS magnetic field ranges from 150 to 1000 Gauss. The lighter green color represents that the height of magnetic field is higher. The symbols are marked the sample positions for measuring the transverse oscillations in Figures~\ref{fig5} and ~\ref{fig6}.} 
\label{fig4}
\end{figure*} 
%

\begin{figure*}
\centering
\includegraphics[width=17cm]{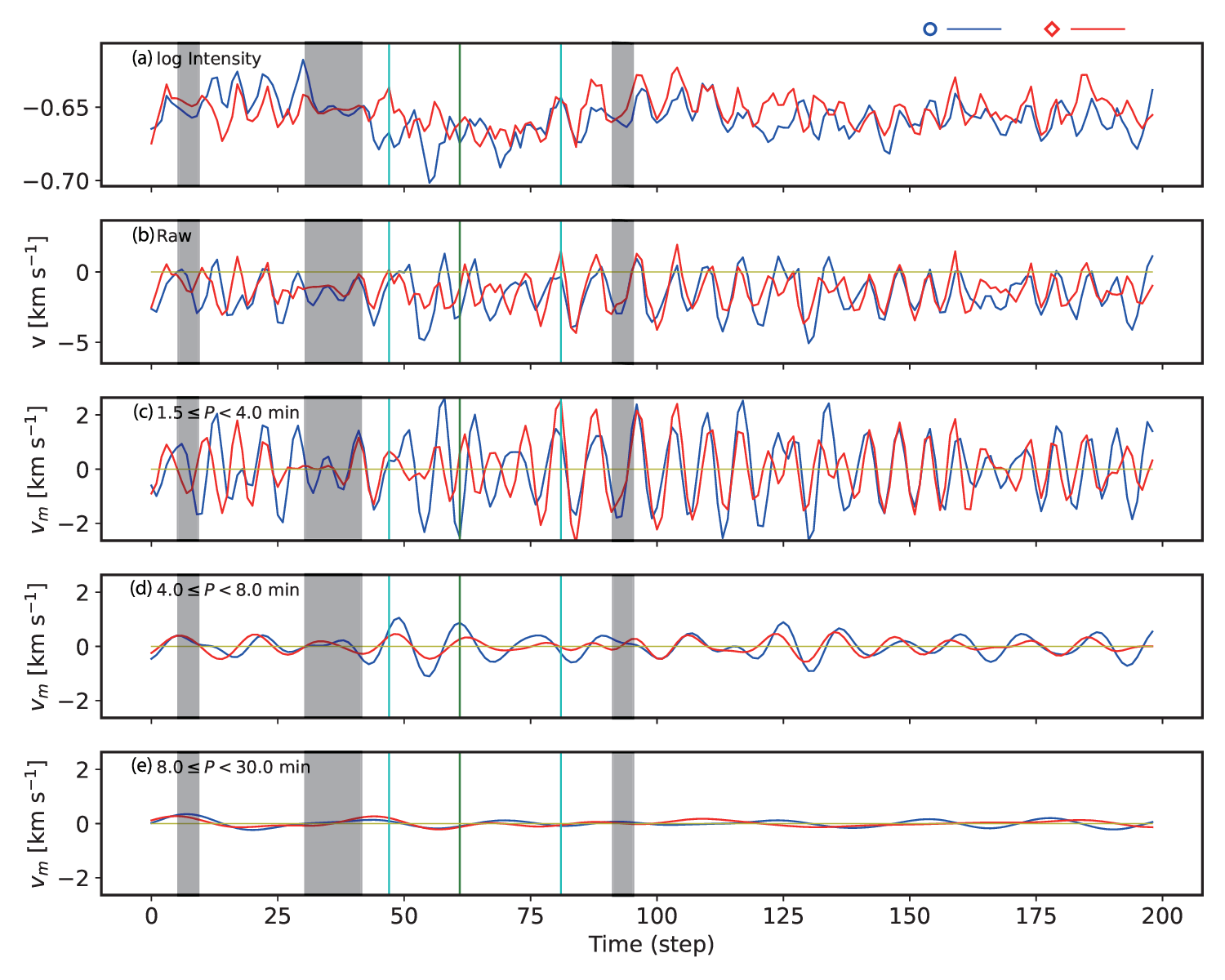} 
\caption{Time variations of (a) H$\alpha$ chromospheric intensity, (b) chromospheric LOS velocity and (c-e) band-filtered LOS velocities at two points, `o' and `$\diamondsuit$' in the umbra, as marked in Figure~\ref{fig4}. The gray shaded regions correspond to the predicted data using an autoregressive model since data are missing. The green vertical line indicates the observing time of the maps presented in Figure ~\ref{fig4}. The two cyan vertical lines indicate the start and end time for the correlation analysis. }
\label{fig5}
\end{figure*}

\begin{figure*}
\centering
\includegraphics[width=17cm]{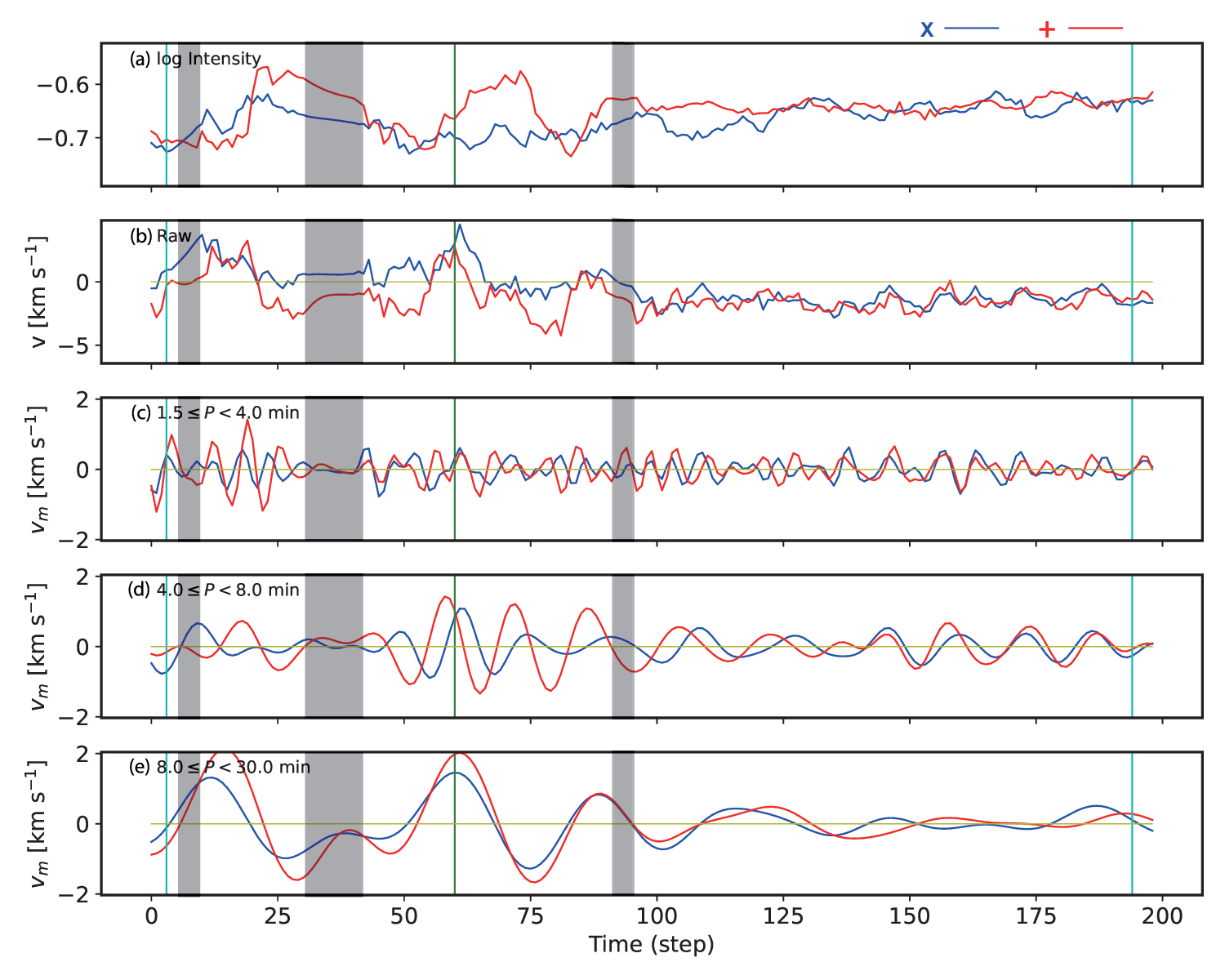} 
\caption{Same as Figure~\ref{fig5}, but for the different two points, 'x' and '+' along the superpenumbral fibril in Figure~\ref{fig4}. }
\label{fig6}
\end{figure*}

\subsection{FISS observations and analysis}
We analyzed the H$\alpha$ imaging spectral data obtained by the FISS to detect transverse MHD waves in the chromosphere. The H$\alpha$ data obtained by using 32$\mu$m slit and the spectral sampling is 0.019 $\mathrm{\AA}$. The raster images are obtained by scanning a slit with a step size of $0.16 \arcsec$, and the spatial sampling along the slit is also $0.16 \arcsec$. In this observing run, the slit scans 120 steps of which FOV is $19.2 \arcsec \times 40 \arcsec$ with a time cadence of 22 seconds. The FISS observes the trailing sunspot region from 20:11:55 to 21:59:03 UT.  The spectral data were calibrated (corrected for the flat field, dark current, and stray light) and compressed by the PCA (Principal Component Analysis) method \citep{chae_etal2013}. During about 1.5 hour observation, there are data gap due to the bad seeing condition in the periods of 20:19:21 - 20:20:44 UT, 20:28:52 - 20:32:45 UT, and 20:50:51 - 20:52:19 UT.  For the analysis of the time series, we predicted and filled data gaps based on observations from preceding and subsequent time points using an autoregressive model.

To derive the physical parameters in the chromosphere, we applied the fast multilayer spectral inversion \citep[MLSI,][]{chae_etal2020, chae_etal2021b, lee_etal2022} to the observed H$\alpha$ spectral profiles. Since the H$\alpha$ spectral profile is formed throughout the low solar atmosphere, with the continua and far wings being formed in the photosphere, and the cores being formed in the upper chromosphere, the MLSI assumes a three-layer atmosphere consisting of the photosphere, lower and upper chromosphere. The inversion provides the estimates of 10 free parameters from the H$\alpha$ spectral profiles, which are four source functions at the boundaries between each layer, two LOS velocities, and two Doppler widths at the boundaries of each layer in the chromosphere, and two parameters for absorption profiles of the photosphere (damping parameter ($a_{p}$) and ratio of peak line absorption to continuum absorption ($\eta$)). 

Among these parameters, we used the upper chromospheric LOS velocity and the chromospheric intensity constructed from the weighted average of the lower and upper chromospheric source functions. Figure~\ref{fig4} displays the H$\alpha$ intensity maps from the source functions of the photosphere, temperature minimum, and chromosphere, as well as the band-filtered chromospheric velocity maps obtained from the MLSI. For the coalignment, SDO/HMI continuum intensity is shown. The contours of HMI LOS magnetogram are also overlaid for outline the umbra and penumbra boundaries. The green lines are the magnetic fields extrapolated by the linear-force-free method. The field lines are aligned with the dark fibrils or velocity strips in the maps of intensity and velocity in the chromosphere. 

For detecting the transverse MHD waves in the chromosphere, we used the same method introduced by \citet{chae_etal2021a, chae_etal2022}. They identified 
transverse MHD waves by detecting LOS velocity oscillations uncorrelated with intensity oscillations along the thread-like structure, which implies the horizontal magnetic field. Since observed oscillations are possibly mixed with the apparently coexisting several waves with different frequencies, we applied several sets of bandpass filtering to velocity and intensity parameter maps. Then, we select two points along the same velocity stripe in the bandpass-filtered velocity maps to detect the wave packets. These velocity stripes typically align or coincide with thread-like intensity structures, which are considered to trace horizontal magnetic field lines. The LOS oscillations along these horizontal field lines represent transverse oscillations. In addition, when we selected velocity stripes, we compared the velocity oscillations to the intensity oscillations at the same positions to confirm that they are not correlated, which implies that the waves are incompressible or very weakly compressible. 

Then, to identify the propagating transverse wave packets, we calculate the cross-correlation coefficient between the two LOS velocities at these selected points. We only define propagating wave packets with cross-correlation values larger than 0.9, and the propagating velocity, velocity amplitude, and periods of the wave packets are also estimated. The direction of wave propagation was determined by analyzing the spatial orientation of the velocity stripe in the bandpass-filtered maps. For consistency, the spatial orientation was always chosen to start from the sunspot center and extend outward. A negative time lag thus indicates inward propagation toward the sunspot center, while a positive time lag represents outward propagation toward the penumbra and superpenumbral region. The details of the correlation analysis methods are described in Section 2.3 of \citet{chae_etal2022}. 

Samples of the selected velocity stripe with two points are presented in Figure~\ref{fig4}: $\times$ and + symbols correspond to the superpenumbral fibril, while $\medcirc$ and $\Diamond$ symbols indicate the selected positions in umbra. The comparison of intensity and velocity oscillations at the positions (panels a and b), and the time variation of bandpass-filtered velocities at two points along the velocity stripe (panels c-e) are shown in Figures~\ref{fig5} and~\ref{fig6}. 

\begin{figure}
\centering
\includegraphics[width=8.5cm]{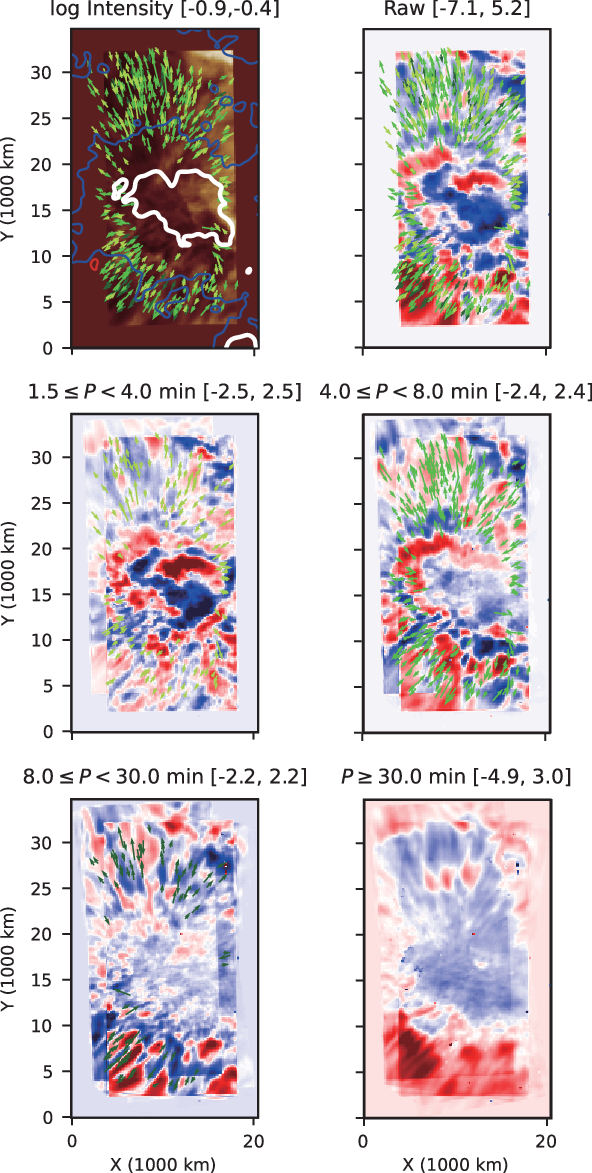}
\caption{Spatial distribution of the 412 detected wave packets during the FISS observing period, 20:11:55 -- 21:59:03 UT. The top panels show the H$\alpha$ intensity and chromospheric LOS velocity maps. Middle and bottom panels display the band-filtered velocity maps. The numeric values in the brackets of each map are the lower and upper bounds for byte-scale display. The arrows depict detected wave packets, with their centers located at the midpoint between two chosen points used for cross-correlation analysis. The direction of the arrows shows the wave propagation direction. The length of the arrows corresponds to the propagation speed, proportional to \(c^{1/3}\). The colors of the arrows distinguish the wave packets from one another based on different period bands: light green for 3-minute waves, medium green for 5-minute waves, and dark green for 10-minute waves. The white, blue, and red contours in the intensity map represent the 1000 Gauss, 150 Gauss, and -150 Gauss, respectively.}  
\label{fig7}
\end{figure}

The identified wave packets are shown as arrows in Figure~\ref{fig7} with intensity and bandpass-filtered LOS velocity images. The direction and length of the arrows indicate the propagating direction and nonlinearly scaled speed of wave packets, respectively. The colors of the arrows represent the different period bands. Using these identified wave packets and determined wave properties, we investigated the statistical characteristics of the propagating transverse waves, which is weakly compressible, in the chromosphere related to the FIP fractionation.

\subsection{SDO observations and analysis}



We used AIA and HMI onboard SDO for context images and coalignment between the observations from Hinode/EIS and FISS (Figure~\ref{fig1}). AIA provides multiple temperature images covering $\log (T/\rm{K})=3.7-7.2$ with a high time cadence of about 12 seconds and a spatial resolution of 1.2$\arcsec$. HMI provides full-Sun LOS magnetograms and continuum images at a spatial resolution of $\sim 1 \arcsec$ and a temporal cadence of 45 seconds. The AIA and HMI data were processed using the standard procedure of {\tt aia\_prep.pro}, which aligns the multiple wavelength images of AIA and HMI. For coalignment between the EIS and FISS, first, we used the processed AIA and HMI observations as the reference image. We aligned the EIS \ion{Fe}{xii} 195.12 $\rm{\AA}$ raster image with the AIA 193 $\rm{\AA}$ filter images showing similar temperature plasma. For the FISS, we aligned AIA 1600 $\rm{\AA}$ and HMI continuum intensity images with raster images from a far wing of H$\alpha$ line which display photospheric intensity. 

We also used HMI LOS magnetograms as an input parameter for the magnetic field extrapolation to investigate the magnetic environment of the sunspot  as well as the magnetic field connection between the chromospheric and coronal structures.
In order to assess the magnetic connection between the chromospheric waves and the coronal plasma composition,
we have extrapolated the magnetic field from the solar surface into the three-dimensional space above by implementing the linear-force-free extrapolation of photospheric magnetic field measurements using the Green's function method \citep{chiu&hilton_1977}. We use the SDO/HMI line-of-sight magnetograms taken at 20:39 UT. The force-free parameter, $\alpha = -0.002$, was selected by evaluating the alignment between the constructed field lines and the EUV loop structures in the SDO/AIA 171 $\rm{\AA}$  image when projected onto the image plane. The extrapolated magnetic field structure is illustrated in Figures~\ref{fig4} and \ref{fig_3dB}. Figure~\ref{fig4} shows the magnetic field configuration in the regions where waves are observed with FISS. The magnetic field lines are depicted in green, with lighter shades representing higher field line heights. Figure~\ref{fig_3dB} demonstrates how these chromospheric magnetic field structures are connected to the corona.

\section{Results}
\subsection{Plasma composition and coronal plasma properties}

Figure~\ref{fig3}(d) presents the map of the FIP bias factor which indicates the plasma composition. A FIP bias factor close to 1 corresponds to photospheric abundance, whereas values exceeding 2-3 indicate coronal abundance, suggesting highly fractionated plasma. 
Comparing the plasma composition map with coronal and photospheric images, we found that the highly fractionated plasma is mainly observed at the footpoints of the brighter compact loops and surrounding the sunspot penumbra, while the fainter long loops and sunspot umbra show unfractionated plasma with a FIP bias factor close to 1-1.5. This agrees with the findings of \citet{baker_etal2021}, which indicated that highly fractionated plasma is linked to coronal loops rooted near the penumbra experiencing Alfv\'enic perturbations, while little to no fractionation is observed in the corona above the core of the umbra. 

In comparison to Figure~\ref{fig2}, which illurstrates properties of coronal temperature plasma, such as intensity, LOS velocity, and nonthermal velocity maps measured from \ion{Fe}{xii} 195.12 $\rm{\AA}$, we find that the highly fractionated plasma is mostly redshifted with a speed of $\sim10~\rm{km~s^{-1}}$ along the bright loop structure. Additionally, some regions of high FIP bias plasma on the eastern side of the trailing sunspot exhibits a blueshifted speed of $\sim10~\rm{km~s^{-1}}$. The regions with high FIP bias factor values display significant nonthermal velocities of $\sim 40-60 ~ \rm{km~s^{-1}}$, consistent with the observations by \citet{baker_etal2013} who reported high nonthermal velocities at loop footpoints associated with highly fractionated plasma.  

\subsection{Spectroscopic detection of transverse MHD waves and their statistical properties}
From the spectroscopic detection method, we detect 412 weakly compressible transverse wave packets in the FISS H$\alpha$ spectral data, which imply that the transverse MHD waves are abundant in the sunspot region. Figure~\ref{fig7} illustrates the detected wave packets during 20:11:55 -- 21:59:03 UT with arrows. From the spatial distribution of the detected wave packets, we found that the transverse waves are dominantly distributed along the penumbral and superpenumbral fibrils starting from the umbra-penumbra boundaries. As explained in Section 2.2, we only identified the LOS velocity oscillations uncorrelated with the intensity oscillations for the weakly compressible transverse waves. Our analysis, which relied on LOS velocity oscillations to detect transverse waves, restricts our ability to measure transverse waves definitively in the umbral region, where the magnetic field lines are vertical. Most LOS velocity oscillations in the umbra correlate with intensity oscillations (Figure~\ref{fig5}), suggesting that the wave packets in the umbra are slow-mode waves. Since our primary focus is on detecting incompressible transverse waves, the observed compressibility of these waves implies that they are not relevant to our study.

To understand the properties of these identified transverse MHD waves in this sunspot, we statistically investigated their period, velocity amplitude, propagation speed, and direction. Figure~\ref{fig8} shows the number distributions of the wave properties. The top, middle, and bottom panels show the histograms of the wave period ($P$), propagating speed ($c$), and velocity amplitude ($V$), respectively. The number distribution of the propagating direction of the waves is also presented on the histogram of the period. 
The mean values and variation range of the wave properties are summarized in Table \ref{tbl-2}. 

The overall statistical properties of the identified transverse waves in this sunspot are similar to the earlier results reported by \citet{chae_etal2022} and \citet{kwak_etal2023}, which have examined the statistical properties of the 
transverse waves in a superpenumbral region around a small sunspot (pore) and fibrils of a quiet network region respectively. 
For all wave packets, the mean values of the period, velocity amplitude, and propagation speed in this sunspot, mainly are about 6 minutes, 0.74 $\rm{km~s^{-1}}$, and 79 $\rm{km~s^{-1}}$, which are comparable to 
the values found from superpeumbral fibrils around the pore by \citet{chae_etal2022}. Below we make detailed comparisons in the statistics between our study and previous ones.

First, the wave periods are classified into three distinct groups based on the detected transverse wave packets. Most wave packets fall within the 5-minute band. Additionally, shorter-period and longer-period waves exhibit peaks around the 2.5-minute and 11-minute bands, respectively. A clear boundary exists between the shorter and longer periods, with a minimum around 7 minutes. 
It is consistent with the 
transverse wave properties in the pore reported by \citet{chae_etal2022}, where two distinct groups of 3-minute and 10-minute waves were identified, also separated around the 7-minute. While the small sunspot pore exhibited a peak in the number distribution at a period of 3 minutes in shorter-period range, the sunspot with a penumbra analyzed in this study shows two distinct peaks: one at 3 minutes and a dominant one at 5 minutes. For the longer-period waves(> 7 minutes), the distribution is broader without showing distinct peaks. The results indicate that the sunspot analyzed in this study also shows the possibility that short-period and long-period waves may have different generation mechanisms as suggested by \citet{chae_etal2022}. In addition, the presence of a penumbra may induce the 5 minute waves, within the typically shorter-period waves, which become dominant in the overall distribution.

Second, we examined the statistics of propagation speed $c$ depending on their wave period groups. The propagation speeds for all waves range from 12 $\rm{km~s^{-1}}$ to 562 $\rm{km~s^{-1}}$. 
The blue, orange, and green colors in the middle and bottom rows of Figure~\ref{fig8} represent each different period group, 2.5-minute, 5-minute, and 11-minute period bands. The mean values of propagation speed for each period band are 57 $\rm{km~s^{-1}}$, 
78 $\rm{km~s^{-1}}$, 
and 104 $\rm{km~s^{-1}}$, 
respectively. The mean propagation speed for all waves is 79 $\rm{km~s^{-1}}$. 
These values are compatible with the propagation speeds in superpenumbral fibrils previously reported by \cite{morton_etal2021} based on the imaging detection as well as 
the speeds reported by \cite{chae_etal2022} using the spectroscopic detection.

Third, we examined the statistics of velocity amplitude $V$ in the same way as above. The mean velocity amplitudes are 0.58 $\rm{km~s^{-1}}$, 
0.81 $\rm{km~s^{-1}}$, 
and 0.76 $\rm{km~s^{-1}}$ 
for 2.5-minute, 5-minute, and 11-minute bands, respectively. All the values range from 0.28 $\rm{km~s^{-1}}$ to 2.53 $\rm{km~s^{-1}}$, which is similar with the reported amplitude of transverse waves in superpenumbral fibrils previously observed \citep{morton_etal2021, chae_etal2022}. They, however, are found to be smaller than those reported from quiet Sun network fibrils  \citep{kwak_etal2023}, roughly by a factor of two.

Last, we find that the distribution of the propagation directions, inward waves and outward waves  are comparable to each other in  number, regardless of the wave period (see top panel of Figure~\ref{fig8}). This result is somewhat different from previous results. In the superpenumbral fibrils in the pore, \citet{chae_etal2022} found that the waves in the 3-minute band is predominantly outward, whereas the waves in the 10-minute band are either inward or outward with equal occurrence rate. 
In the quiet Sun fibrils, meanwhile, \citet{kwak_etal2023} reported that most waves propagate outward regardless of period, even in the 10-minute band.

\begin{table*}
\caption{Mean and standard deviations of physical parameters of detected transverse waves}
\label{tbl-2}
\centering
\begin{tabular}{c c c c c}
\hline\hline
Parameters & 2.5-minute waves & 5-minute waves & 11-minute waves & All waves\\
\hline 
    Number of wave packets & 95 & 222 & 95 & 412 \\
    $P$, period (minutes) & 2.45 $\pm$ 0.32 & 4.83 $\pm$ 0.78 & 11.46 $\pm$ 3.18  & 5.81 $\pm$ 3.63  \\
    $c$, propagation speed ($\rm{km~s^{-1}}$) & 57.02 $\pm 69.86 $ & 77.97 $\pm$ 85.37 & 104.42 $\pm$ 109.66 & 79.24 $\pm$ 89.92 \\ 
    $V$, velocity amplitude ($\rm{km~s^{-1}}$) & 0.59 $\pm$ 0.29 & 0.81 $\pm$ 0.44 & 0.76 $\pm$ 0.34 & 0.74 $\pm$  0.39 \\
    Inward propagation fraction (\%) 	& 46 & 42 & 44 & 43 \\
\hline
\end{tabular}
\end{table*}

\begin{figure*}
\centering
\includegraphics[width=17cm]{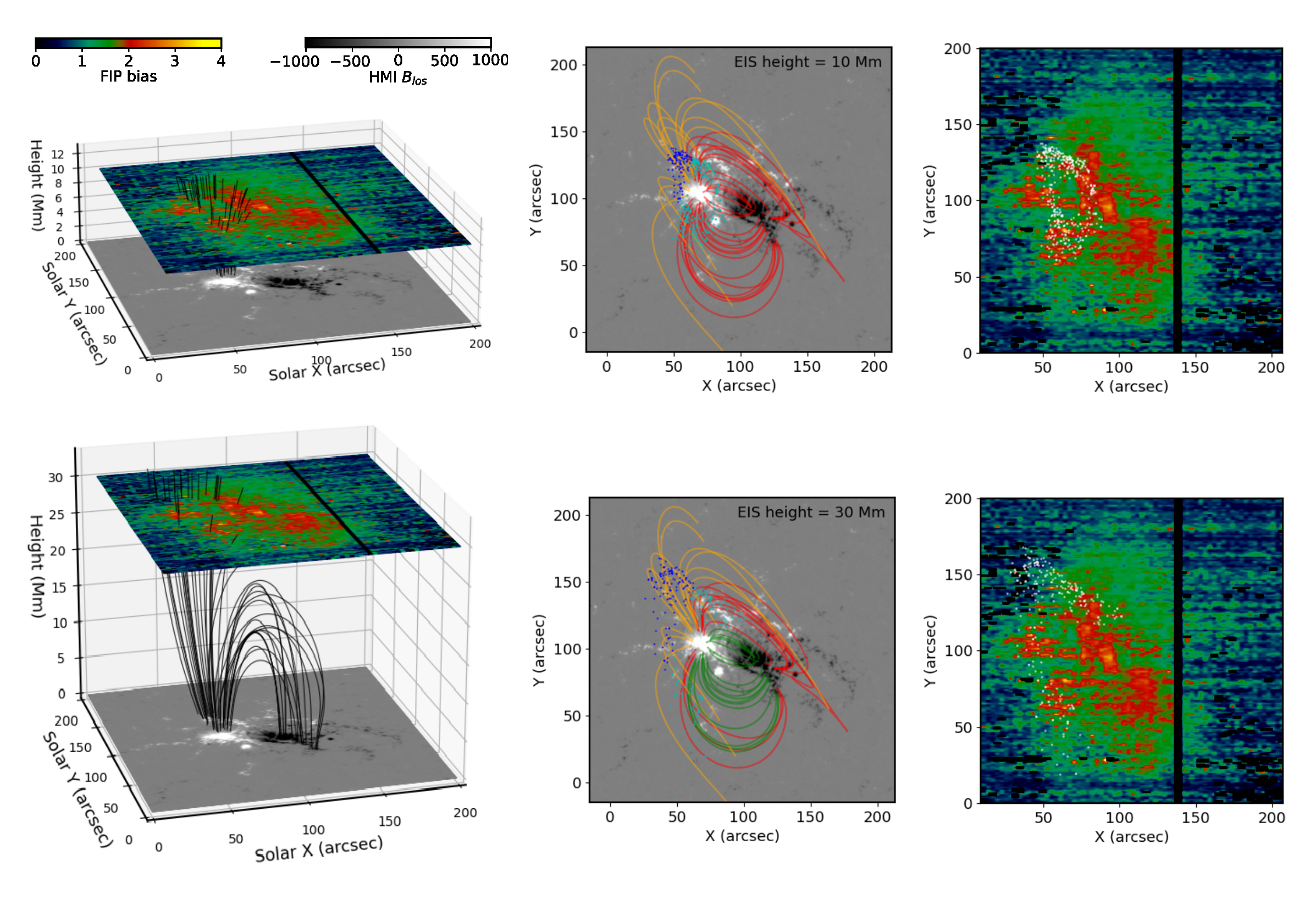} 
\caption{Left panels: 3D view of extrapolated magnetic fields from the HMI magnetogram using a linear force-free field model with $\alpha = -0.002$. The fields connect to the EIS FIP bias map supposedly at heights of 10 Mm (top) and 30 Mm (bottom), respectively. The magnetic field lines originate from the wave packet locations in the chromosphere at an assumed height of 2 Mm. Middle panels: Projected magnetic fields with yellow indicating open fields. Red and green lines represent closed field lines, with red indicating heights above the EIS FIP bias map and green below. Blue and cyan dots mark where the field lines connect to the EIS map for open and closed fields, respectively. Right panels: Connected regions (white dots) projected onto the EIS FIP bias map at different heights.} 
\label{fig_3dB}
\end{figure*}

\begin{figure}
\centering
\resizebox{\hsize}{!}{\includegraphics{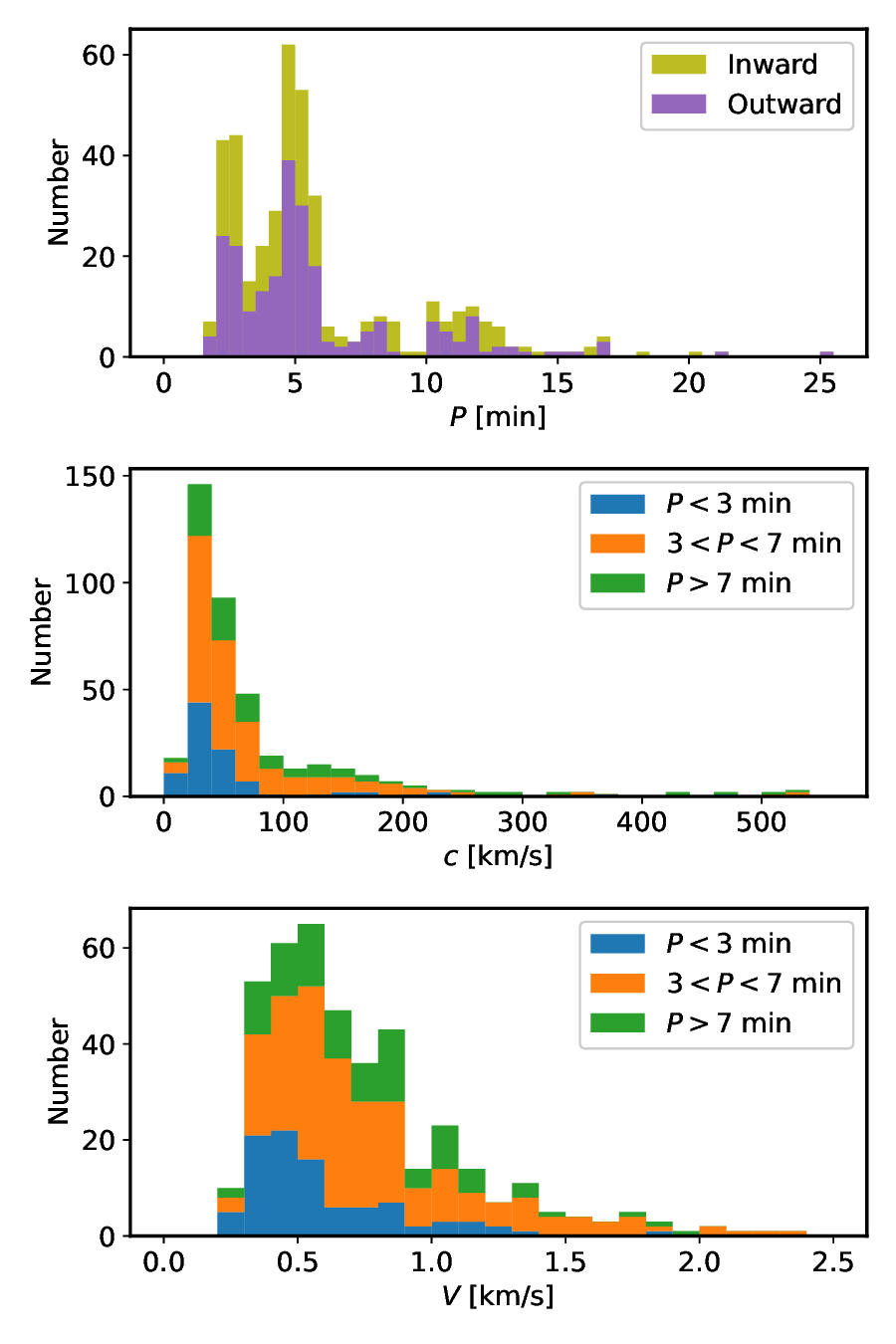}}
\caption{Stacked histograms of (top) period ($P$), (middle) propagation speed ($c$), and (bottom) velocity amplitude ($V$). Period values are categorized into outward and inward waves, while propagation speed and velocity amplitude are grouped into three period bands: 2.5-minute, 5-minute, and 11-minute waves.} \label{fig8}
\end{figure}

\subsection{Connection of the coronal plasma composition and chromospheric transverse MHD waves}

To examine whether the chromospheric transverse waves contributes to the coronal plasma composition, we investigated the relation between the spatial distributions of the transverse waves and the FIP bias. In Figure~\ref{fig3}(d), the highly fractionated plasma is concentrated along the bright loop structures and around the penumbral region, but not within the umbra, as indicated by the black contours. This is consistent with our observation that transverse waves in the chromosphere are found from the umbra-penumbra boundary to the superpenumbral region but not within the umbra itself.

Among the extrapolated three-dimensional magnetic fields, we display the field lines originating from the region where the transverse waves were detected in the chromosphere, as shown in Figure~\ref{fig_3dB}. We suppose the chromospheric height observed in the H$\alpha$ core line to be approximately 2 Mm. 
Starting with 412 detected chromospheric wave packets, we traced the extrapolated magnetic fields to identify connections to the EIS FIP bias map at heights of 10 Mm (top panel) and 30 Mm (bottom panel). 

When projecting the magnetic field lines, the central region tends to be dominated by field lines located at lower heights. 
Considering the optically thin nature of coronal emissions, which are integrated along the LOS across various heights, the observed emissions are likely to include contributions from lower altitudes. This is the reason why we examine the connections at different heights. 

\begin{figure}
\centering
\resizebox{\hsize}{!}{\includegraphics{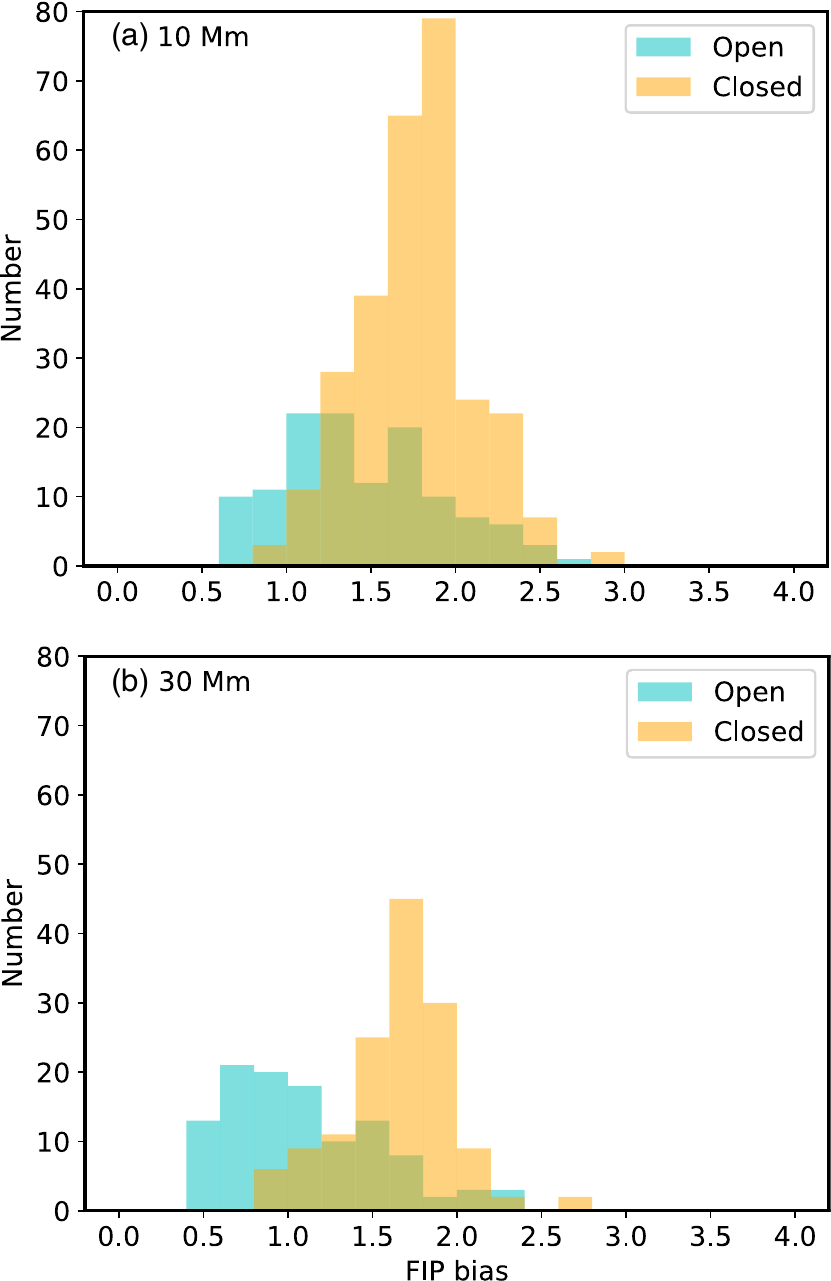}} 
\caption{Histograms of FIP bias values from regions linked to extrapolated magnetic fields originating from 412 chromospheric wave packets, at assumed coronal heights where FIP bias is observed: (a) 10 Mm and (b) 30 Mm. The distributions are based on 404 intersecting regions (open: 124, closed: 280) at 10 Mm and 250 regions (open: 111, closed: 139) at 30 Mm reflecting height-dependent connectivity. 
Blue represents regions connected through open fields, while yellow indicates those connected through closed fields.} \label{fig10}
\end{figure}

By comparing the detected transverse waves with the FIP bias map, linked to the extrapolated magnetic fields from these regions, we found that FIP bias values exhibit different trends depending on whether the regions are connected via open or closed magnetic fields. 
As shown in Figure~\ref{fig10}, closed magnetic fields correspond to higher FIP bias values, with mean values decreasing slightly from 1.8 at 10 Mm to 1.7 at 30 Mm, indicating continued plasma fractionation with altitude. In contrast, plasma fractionation in regions with open magnetic fields diminishes with height, with mean FIP bias values decreasing from 1.5 at 10 Mm to 1.1 at 30 Mm, suggesting little to no fractionation at greater altitudes. 
This observation aligns well with Laming's ponderomotive force model \citep{laming2015, laming2017}, which predicts enhanced fractionation due to ponderomotive forces driven by Alfv\'en waves, particularly in closed loops where resonance effects amplify these forces, whereas open field lines exhibit little resonance and thus reduced fractionation.

\section{Discussion and Summary}

As presented in the previous section, the transverse waves detected along the sunspot penumbra in the chromosphere are associated with fractionated coronal plasma characterized by a high FIP bias. We have examined the statistical properties of these propagating waves, including their periods, velocity amplitudes, propagation speeds, and directions, using a spectroscopic detection method (Section 3.2). 

We interpreted the observed wave properties using an analytical solution for Alfv\'en waves in a medium where the Alfv\'en speed varies along magnetic field lines with a constant gradient, as proposed by \citet{chae&lee_2023}, and in the context of the ponderomotive force model for FIP fractionation described by \citet{laming2015, laming2017}. Notably, the solution and model suggest that downward-propagating low-frequency Alfv\'en waves, with periods of a few minutes originating from the corona, drive significant FIP fractionation near the top of the chromosphere. 

First, to assess the frequency characteristics of the detected waves, we examined the dominant wave periods, which ranged from 1.5 to 25 minutes, peaking around 5 minutes. The 22-second cadence of FISS observations limited detection to periods longer than 1 minute. Most detected waves thus fell within the low-frequency range, as confirmed by the Alfv\'enic cutoff frequency $\omega_{c}$ derived from \citet{chae&lee_2023}, 
\begin{equation}
\omega_{c}=\frac{1}{2} \frac{dv_{A}}{ds} =\frac{1}{2} \frac{v_{A, 1}-v_{A, 0}}{s_{1}-s_{0}} 
\end{equation}
where $s_{0}$ and $s_{1}$ represent the heights of the layers, and $v_{A, 0}$ and $v_{A, 1}$ are the Alfv\'en speeds at those heights. The Alfv\'en speed is given by $v_{A}=B/\sqrt{4 \pi \rho}$, where $B$ is the magnetic field strength and $\rho$ is the mass density. The magnetic field strength at each height is obtained from an extrapolated linear-force-free magnetic field model based on HMI photospheric magnetic field measurements, averaged over the penumbral region.The chromospheric density is taken from the FAL-P model \citep{fontenla_etal1993}, while the coronal density is measured using the observed EIS \ion{Fe}{xiii} line pairs in the high-FIP region, ranging from $10^{9}$ to $10^{9.3} ~\mathrm{cm^{-3}}$.The parameters and estimated cutoff frequency are summarized in Table \ref{tbl-3}. The estimated cutoff frequencies, $\omega_{c}$, range from 0.5 to $0.05~\mathrm{rad ~s^{-1}}$ at loop heights of 10 - 30 Mm. 
The corresponding cutoff periods, $P_{c}$, are approximately 13 - 126 seconds. Thus, the detected transverse waves with periods of several minutes for these loops are categorized as evanescent low-frequency waves.

Second, we examined the propagation direction of the detected transverse wave. The transverse waves in this sunspot penumbra and super-penumbral fibrils exhibit comparable proportions of inward and outward waves. In contrast, quiet network regions, where little to no FIP effect is typically observed, predominantly exhibit outward-propagating waves along the fibrils \citep{kwak_etal2023}. These findings indicate that revealing the propagation direction is crucial because it determines how these waves contribute to FIP fractionation. 

Note we interpret the observed inward waves as the chromospheric manifestation of downward-propagating waves from the corona. The extrapolated magnetic field configuration, shown in Figures~\ref{fig4} and \ref{fig_3dB} supports our interpretation. The magnetic field lines in the sunspot region become increasingly inclined outward and upward from the center toward the penumbra. Based on this configuration, outward waves from the sunspot center to the penumbra were interpreted as upward-propagating waves, and inward waves directed toward the center were associated with downward-propagating waves.

While downward-propagating waves are expected to be dominant in high FIP bias regions according to the ponderomotive force model \citep{laming2015, laming2017, chae&lee_2023}, our observations show comparable proportions of upward- and downward-propagating waves. However, this does not mean that downward waves are less prevalent. Compared to quiet network fibrils, where low-FIP regions predominantly exhibit upward-propagating waves, downward waves appear more prominent in high-FIP bias regions, suggesting their significant role in FIP fractionation. Additionally, while upward waves may not directly induce FIP fractionation, their reflection contributes to the generation of additional downward waves, which may influence the fractionation process.

Regarding the source of inward (downward)-propagating waves, which are important for FIP fractionation, both coronal and photospheric origins are plausible. Coronal waves, such as those generated by magnetic reconnection or nanoflares, may travel downward to loop footpoints at the top of the chromosphere. These waves could resonate with loops, causing significant fractionation near the upper chromosphere or the transition region as proposed by the model of \citet{laming2017}. Another possible source is the photospheric waves that reflect at the transition region due to steep gradients in density and magnetic field, creating resonant or nonresonant waves. The observed portion of upward waves may indicate photospheric-originated transverse MHD waves. Detecting wave reflection directly may be helpful for validating the origin of the observed low-frequency downward waves, which are important for FIP fractionation. However, our observations were limited by the lack of multi-wavelength spectroscopic data covering different atmospheric heights. Even though wave reflection itself was not detected, the observed balance of inward and outward waves suggests a dynamic sunspot penumbra influenced by contributions from both coronal and photospheric waves. Therefore, we consider the FIP fractionation may be generated by a mix of coronal and photospheric originated transverse MHD waves, and the degree of the fractionation would be determined by their wave amplitude and degree to which the wave is resonant with the loop as suggested by \citet{laming2019}, \citet{Mihailescu_2023}, and \citet{Murabito_etal2024}.

Third, the resonant/nonresonant wave characteristics were examined. The resonant wave with a coronal loop refers the condition where the wave travel time from one loop footpoint to the other is an integral number of wave half periods. Recent studies by \citet{Mihailescu_2023} and \citet{Murabito_etal2024} suggested that the characteristics of the resonant/nonresonant wave is also crucial parameter for the fractionation process, which determine where the ponderomotive force is generated. Resonant waves drive the ponderomotive force close to the top of the chromosphere producing mild fractionation because of the ionized background gas, while nonresonant waves drive the ponderomotive force at lower heights of the chromosphere, which results in stronger fractionation levels because the background gas is neutral. To examine the measured transverse waves are resonant or nonresonant with loops, we estimate the resonant frequency, $f_{\mathrm{resonance}}=v_{A}/2L$, where $v_{A}$ is the Alfv\'en speed and $L$ is the loop length. With the loop lengths of 20-200 Mm and magnetic field strengths of 10-300 G based on the extrapolated magnetic field model, and coronal density measured from EIS \ion{Fe}{xiii}, the estimated resonant frequencies for the loops are 0.3-0.003 $\mathrm{s^{-1}}$, which correspond to the periods of approximately 3-300 s. Resonant frequencies in shorter loops ($L$ <100 Mm) are much higher than the frequencies of the observed chromospheric transverse waves, implying these waves are nonresonant. For longer loops ($L$ >100 Mm) the resonant frequencies fall within the range of our observations, suggesting potential resonance. We find that the higher FIP biases observed in low-height loops likely result from nonresonant waves, in line with the results of \citet{Murabito_etal2024}.

The observed properties of transverse waves in our study, which highlight the prominence of downward-propagating low-frequency waves, are consistent with the predictions of the ponderomotive acceleration model. However, several issues remain unresolved. First, there might be a suspicion that the waves we detected are not transverse waves, but the projection of longitudinal waves of magnetoacoustic type \citep{chelpanov&kobanov_2024}. In order to resolve this ambiguity, we have tried our best by including in the analysis only the waves where velocity and intensity are not significantly correlated. Nevertheless, we admit that we can not fully exclude the possible mixture of transverse and longitudinal waves. The clearer resolution of this issue needs more sophisticated data and analysis in the future.  

Second, the measured wave velocity amplitudes appear insufficient to produce significant FIP fractionation. 
As ponderomotive acceleration is proportional to the square of the wave amplitude, the observed wave velocities, ranging from 0.28 to 2.53 $\mathrm{km~s^{-1}}$ with a mean value of 0.7 $\mathrm{km~s^{-1}}$ are too low to achieve the required acceleration. Based on an analytical approach, \citet{chae&lee_2023} assessed ponderomotive acceleration at the top of the chromosphere, where FIP fractionation is believed to predominantly occur. In their analysis, low-frequency Alfv\'en waves propagating downward were identified as significant, producing an upward-directed ponderomotive force proportional to $V^{2}/4\vert H_{D} \vert$. They indicated that velocity amplitudes of approximately 4 $\mathrm{km~s^{-1}}$ could achieve ponderomotive acceleration exceeding gravitational acceleration by an order of magnitude. For a ponderomotive acceleration comparable to gravitational acceleration, velocity amplitudes over 1.4 $\mathrm{km~s^{-1}}$ are required. 
In this regard, it is recalled that our spectroscopic method measures the amplitude of velocity oscillation in the vertical direction. We think that there is a good possibility that the waves may be oscillating in the horizontal direction as well, probably with larger amplitudes. The recent observations (Bate et al. \citeyear{bate_etal2024}; Kwak et al. in prep.) supported this possibility, and suggested the horizontal-to-vertical velocity amplitude ratio is larger than two. If this is the case, the vertical velocity amplitude of 0.7 $\mathrm{km~s^{-1}}$, in fact, may be an indication of the total velocity amplitude larger than 1.5 $\mathrm{km~s^{-1}}$. 
Further studies are needed to address whether current observational techniques miss the full spectrum of wave activity, or if physical processes inherently limit wave amplitudes.

Finally, high-frequency waves with periods shorter than 1 minute may not be detectable due to limitations in current observations. The resonant frequencies with shorter coronal loops are approximately 0.3-0.02 $\mathrm{s^{-1}}$, corresponding to periods shorter than 1 minute. High-cadence observations, such as sit-and-stare modes or future spectroscopic missions (e.g., Solar-C), would be beneficial for further research. Moreover, to investigate the resonant/nonresonant effects and to gain more precise insights into the wave propagation directions and the role of wave reflection, spectroscopic observations of spectral lines formed at different atmospheric heights and involving various elements across different solar structures are required.

In summary, this study explores the relationship between chromospheric transverse waves and coronal plasma composition using simultaneous GST/FISS and Hinode/EIS observations. Previously, only a few observations of Alfv\'enic perturbations associated with coronal fractionated plasma have been reported through the integration of spectropolarimetric observations of the solar chromosphere and spectroscopic observations of the solar corona \citep{baker_etal2021, stangalini_etal2021}. Our study provides further observational evidence, showing that chromospheric transverse MHD waves detected in the sunspot penumbra are associated with fractionated coronal plasma exhibiting a high FIP bias, along with detailed wave properties.
Our observations support the ponderomotive force model, which proposes that MHD waves play a crucial role in generating the FIP effect \citep{laming2015, laming2017, laming2019}. Furthermore, the properties of chromospheric waves and coronal loop characteristics could provide essential observational constraints for understanding the FIP fractionation process. 

\begin{table}
\caption{Estimated Alfv\'enic cutoff frequencies based on loop parameters}
\label{tbl-3}
\centering
\begin{tabular}{c c c c}
\hline\hline
Parameters & $s_{0}$ & \multicolumn{2}{c}{$s_{1}$}\\
 & (km) & \multicolumn{2}{c}{(km)}\\
 \cline{3-4} \\
  & 500 & $1 \times 10^{4}$ & $3 \times 10^{4}$  \\
\hline 
    $B$ (G) & 700 &  200 & 50  \\
    $\rho$ ($\mathrm{g~ cm^{-3}}$) & $2.8 \times 10^{-11}$ & $3.3 \times 10^{-15}$  & $1.7 \times 10^{-15}$ \\
    $v_{A}$ ($\mathrm{km~s^{-1}}$) & $3.7 \times 10^{2}$ & $1 \times 10^{4}$ & $3.4 \times 10^{3}$  \\ 
    $\omega_{c}$ ($\mathrm{rad~s^{-1}}$) & ... &  0.5 & 0.05  \\
    $P_{c}$ (s) & ... & 13 & 126 \\
\hline
\end{tabular}
\end{table}

\begin{acknowledgements}
      This research was supported by the National Research Foundation of Korea (NRF--2021R1A2C1010881, RS--2023--00208117) funded by the Korean Ministry of Science and ICT.  H.K. was supported by Basic Science Research Program through the NRF funded by the Ministry of Education (RS-2024-00452856). E.-K.L. was supported by the Korea Astronomy and Space Science Institute under the R\&D program of the Korean government (MSIT; No. 2025-1-850-02). The data are courtesy of the science teams of FISS/GST, Hinode, and SDO. We gratefully acknowledge the use of data from the Goode Solar Telescope (GST) of the Big Bear Solar Observatory (BBSO). BBSO operation is supported by US NSF AGS--2309939 and AGS--1821294 grants and New Jersey Institute of Technology. GST operation is partly supported by the Korea Astronomy and Space Science Institute and the Seoul National University. Hinode is a Japanese mission developed and launched by ISAS/JAXA, with NAOJ as domestic partner and NASA and UKSA as international partners. It is operated by these agencies in co-operation with ESA and NSC (Norway). HMI and AIA are instruments on board SDO, a mission for NASA's Living With a Star program.
\end{acknowledgements}

%
%

\bibliographystyle{aa}
\bibliography{fip_alfvenwave}

\end{document}